%% file: jones_sardinia.tex
\documentclass[
    ,final            % use final for the camera ready runs
  ]
  {aipproc}

\layoutstyle{6x9}

\begin{document}

\title{Observations of Radio Giant Pulses with GAVRT}

\classification{97.60.Gb, 95.55.Jz, 95.85.Bh}
\keywords      {Crab pulsar, Giant pulses, Radio astronomy instrumentation}

\author{Glenn Jones}{
  address={Jansky Postdoctoral Fellow, National Radio Astronomy Observatory, USA}
  ,altaddress={California Institute of Technology, Pasadena, CA}
}

\include{aas_macros}

\begin{abstract}
Radio giant pulses provide a unique opportunity to study the pulsar radio emission mechanism in exquisite detail. Previous studies have revealed a wide range of properties and phenomena, including extraordinarily high brightness temperatures, sub-nanosecond emission features, and banded dynamic spectra. New measurements of giant pulse characteristics can help guide and test theoretical emission models. To this end, an extensive observation campaign has begun which will provide more than 500 hours on the Crab with a 34-meter antenna located in California, USA. The observations are being done as part of an educational outreach program called the Goldstone-Apple Valley Radio Telescope (GAVRT). This antenna has a novel wide bandwidth receiver which provides up to 8 GHz of instantaneous bandwidth in the range of 2.5 to 14 GHz. These observations will provide detailed information about the variability, amplitude distribution, and detailed frequency structure of radio giant pulses. In addition, a database of pulses from these observations and others of the Crab pulsar is being created which will simplify multiwavelength correlation analysis.
\end{abstract}

\maketitle

%%%%%%%%%%%%%%%%%%%%%%%%%%%%%%%%%%%%%%%%%%%%
%% MAINMATTER
%%%%%%%%%%%%%%%%%%%%%%%%%%%%%%%%%%%%%%%%%%%%

\section{Introduction}
Giant radio pulses from the Crab pulsar have been extensively studied over the years. Recent studies have largely fallen into two categories. The first are statistical studies of a large number of pulses, in which continuous time series from the telescope are recorded to disk and analyzed offline to extract the giant pulses \cite[e.g.][]{ps07, btk08, ksv10}. The time resolution for these studies has typically been longer than 1 microsecond and bandwidths are typically 100--200 MHz. Most of the available data has been taken around 1400 MHz. Cordes et al. \cite{cbh+04} analyzed giant pulse statistics at several frequencies between 430 to 8000 MHz, but each frequency was observed separately. A few studies have used coordinated observations with multiple telescopes to study giant pulses at widely separated frequencies simultaneously \cite{sbh+99, pku+06}.

The second type of observations have probed the structure of individual pulses themselves, by recording the widest bandwidths technically feasible \cite{hkw+03,he07,jpk+10}. Continuously recording such large bandwidths is impractical, so the recording system must be triggered by a pulse detected in real time. The effective duty cycle and limited telescope time of these experiments has limited the sample sizes to of the order 100 pulses. These studies have uncovered a wealth of structure which has inspired and challenged many theoretical emission models \cite{he07}.

These two categories of observations have raised many questions about the nature of the giant pulse emission mechanism and the effects of the interstellar medium on the radiation. A long term study of giant pulses from the Crab pulsar with higher time resolution and wide instantaneous frequency coverage has been undertaken with the DSS-28 telescope, which will provide new information to help resolve these questions.

\section{The DSS-28 system}
The 34-meter DSS-28 radio telescope is operated by the Jet Propulsion Laboratory and the Lewis Center for Educational Research as part of the GAVRT educational outreach program which aims to enrich K-12 curricula by teaching students about radio astronomy. The students take data with the telescope to help scientists involved in the program. The antenna is equipped with a novel wide bandwidth feed which covers 2.5--14 GHz simultaneously \cite{imb07,jon10}. A set of four independently tunable dual polarization 2 GHz bandwidth receivers provide access to up to 8 GHz of instantaneous bandwidth. A flexible digital signal processing system made up of FPGA boards from the CASPER\footnote{\url{http://casper.berkeley.edu}} group and high performance graphics processing unit (GPU) computer nodes enables a wide variety of pulsar observation modes. The two modes most commonly used for giant pulse observations are described below.

\subsection{Wide bandwidth observations}
In wide bandwidth mode, each 1 GHz subband is digitized and streams into a circular buffer. A trigger signal is formed by incoherently dedispersing the band in real time using the known dispersion measure of the pulsar. Depending on the expected length of the dispersed signal, the available memory can be broken up into many circular buffers. This allows bursts of pulses to be captured which are then transfered to disk for offline processing. Currently, a total of 8 GHz is brought from the receiver system at the vertex of the antenna to the signal processor in the pedestal. Thus, wide bandwidth observations are limited to 8 GHz with a single polarization or 4 GHz with both polarizations. Initial observations have been made with 8 GHz from a single polarization.

\subsection{Continuously recorded observations}
To record every giant pulse from the pulsar, it is possible to send raw voltage data from the digital system to a computer with a high performance GPU for real time coherent dedispersion. The implementation used at DSS-28 is based on the NRAO GUPPI machine \cite{drd+08}. Each GPU node is capable of processing a dual polarization 128 MHz subband. Currently we have two such nodes, so two arbitrarily tuned subbands can be processed simultaneously. The GUPPI code has been modified to search the coherently dedispersed time series on the GPU in real time for giant pulses. When a giant pulse is detected this way, the current block of dedispersed data can be written to disk at full time resolution. The highest detection rate for the coherently dedispersed 128 MHz subband has been found to be at the low end of the frequency range, around 2.6 GHz.

\section{Initial Results and Conclusions}
\begin{figure}
	\label{widepulse}
  \includegraphics[height=.35\textheight]{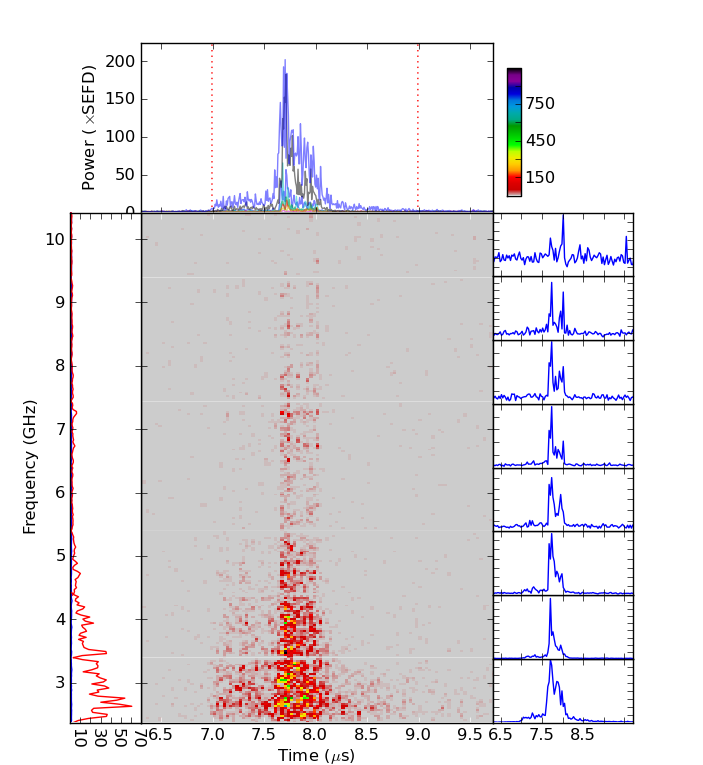}
  \caption{Example of a giant pulse captured at DSS-28. Note the 8 GHz instantaneous bandwidth. The panel to the left of the dynamic spectrum shows the average spectrum of the on-pulse region relative to the off-pulse region. The top and right panels show the intensity time series for each of the 8 subbands. In the top panel, each subband is plotted in a different color. The vertical dotted lines show the on-pulse region. The vertical scale of each of the right panels has been automatically set for each subband. This pulse occured at the rotational phase of the main pulse.}
\end{figure}

The wide bandwidth observations provide the most complete picture of giant pulse emission from 2.5 to 10.5 GHz. An example pulse is shown in figure \ref{widepulse}. The emission clearly spans the entire frequency range. Note that since only one linear polarization is recorded, changes in polarization angle versus frequency could manifest as changes in intensity versus frequency. Faraday rotation in the interstellar medium will cause a change in polarization angle versus frequency, but using a rotation measure for the Crab pulsar of $-42.3$ rad $\mathrm{m}^{-2}$ \cite{mht+05}, we expect intensity variations of less than 30\% across the 2.5 to 10.5 GHz band. Since the rotation angle scales as $\lambda^2$, the effect will be concentrated at the low end of the band. The initial sample of giant pulses shows a wide range of morphological features. As statistical information is accumulated, these features may shed new information on the emission mechanism. The statistics of the spectral characteristics of the pulses should help disentangle propagation effects from the intrinsic properties of the emission. 

One aspect of the GAVRT giant pulse observation campaign is looking for correlation between radio giant pulses and gamma-ray photons detected by the Fermi telescope. For this project, over 100 hours of observations have been made in the continuous recording mode at 2.6 GHz in order to obtain the largest number of pulses possible. The number of giant pulses detected varies dramatically day to day. On some days, a giant pulse is detected every few seconds, while on other days the rate drops below one per every hundred seconds. 
%%%%%%%%%%%%%%%%%%%%%%%%%%%%%%%%%%%%%%%%%%%%
%% Sample figure:
%%
%% The option [height=...] scales the picture to the given height,
%% without it it would be printed at its nominal size
%%%%%%%%%%%%%%%%%%%%%%%%%%%%%%%%%%%%%%%%%%%%

The extremely wide bandwidth receiver and signal processor installed on the DSS-28 antenna are well suited for studying giant radio pulses from the Crab pulsar. The ongoing observation campaign provides much greater simultaneous frequency coverage and observing time than previous studies. The data from this campaign is being stored in a database which will soon be openly accessible to other researchers and will also include data from other telescopes and wavelengths. The data set will allow statistical studies of pulse characteristics, both intrinsic and due to the interstellar medium. 

%%%%%%%%%%%%%%%%%%%%%%%%%%%%%%%%%%%%%%%%%%%%%%%%
%% BACKMATTER
%%%%%%%%%%%%%%%%%%%%%%%%%%%%%%%%%%%%%%%%%%%%%%%%

\begin{theacknowledgments}
Ryan Shannon and Jim Cordes are co-investigators on the GAVRT/Fermi gamma-ray/radio correlation study which is partially supported by NASA through a grant from the Fermi Space Telescope guest observer program. These observations would not be possible without the excellent work by the entire DSS-28 implementation team at Caltech, JPL, and the Lewis Center. The digital signal processor was made possible by generous hardware and software donations from Xilinx, Inc. The author is supported by a Jansky Fellowship from the National Radio Astronomy Observatory which is operated by cooperative agreement between the National Science Foundation and Associated Universities Incorporated.
\end{theacknowledgments}

%%%%%%%%%%%%%%%%%%%%%%%%%%%%%%%%%%%%%%%%%%%%%%%%
%% The bibliography can be prepared using the BibTeX program or
%% manually.
%%
%% The code below assumes that BibTeX is used.  If the bibliography is
%% produced without BibTeX comment out the following lines and see the
%% aipguide.pdf for further information.
%%
%% For your convenience a manually coded example is appended
%% after the \end{document}
%%%%%%%%%%%%%%%%%%%%%%%%%%%%%%%%%%%%%%%%%%%%%%%%

%%%%%%%%%%%%%%%%%%%%%%%%%%%%%%%%%%%%%%%%%%%%%%%%
%% You may have to change the BibTeX style below, depending on your
%% setup or preferences.
%%
%%
%% For The AIP proceedings layouts use either
%%%%%%%%%%%%%%%%%%%%%%%%%%%%%%%%%%%%%%%%%%%%

\bibliographystyle{aipproc}   % if natbib is available
%\bibliographystyle{aipprocl} % if natbib is missing

%%%%%%%%%%%%%%%%%%%%%%%%%%%%%%%%%%%%%%%%%%%
%% You probably want to use your own bibtex database here
%%%%%%%%%%%%%%%%%%%%%%%%%%%%%%%%%%%%%%%%%%%
\bibliography{sardinia}

%%%%%%%%%%%%%%%%%%%%%%%%%%%%%%%%%%%%%%%%%%%
%% Just a reminder that you may have to run bibtex
%% All of it up to \end{document} can be removed
%% if you don't like the warning.
%%%%%%%%%%%%%%%%%%%%%%%%%%%%%%%%%%%%%%%%%%%
\IfFileExists{\jobname.bbl}{}
 {\typeout{}
  \typeout{******************************************}
  \typeout{** Please run "bibtex \jobname" to optain}
  \typeout{** the bibliography and then re-run LaTeX}
  \typeout{** twice to fix the references!}
  \typeout{******************************************}
  \typeout{}
 }

\end{document}

%% file: aas_macros.tex
%
%  These Macros are taken from the AAS TeX macro package version 4.0.
%  Include this file in your LaTeX source only if you are not using
%  the AAS TeX macro package and need to resolve the macro definitions
%  in the BibTeX entries returned by the ADS abstract service.
%
%  If you plan not to use this file to resolve the journal macros
%  rather than the whole AAS TeX macro package, you should save the
%  file as ``aas_macros.sty'' and then include it in your paper by
%  using a construct such as:
%	\documentstyle[11pt,aas_macros]{article}
%
%  For more information on the AASTeX macro package, please see the URL
%	http://www.aas.org/publications/aastex.html
%  For more information about ADS abstract server, please see the URL
%	http://adswww.harvard.edu/ads_abstracts.html
%

% Abbreviations for journals.  The object here is to provide authors
% with convenient shorthands for the most "popular" (often-cited)
% journals; the author can use these markup tags without being concerned
% about the exact form of the journal abbreviation, or its formatting.
% It is up to the keeper of the macros to make sure the macros expand
% to the proper text.  If macro package writers agree to all use the
% same TeX command name, authors only have to remember one thing, and
% the style file will take care of editorial preferences.  This also
% applies when a single journal decides to revamp its abbreviating
% scheme, as happened with the ApJ (Abt 1991).

%\def#1{{#1}}

\def\aj{{AJ}}                   % Astronomical Journal
\def\araa{{ARA\&A}}             % Annual Review of Astron and Astrophys
\def\apj{{ApJ}}                 % Astrophysical Journal
\def\apjl{{ApJ}}                % Astrophysical Journal, Letters
\def\apjs{{ApJS}}               % Astrophysical Journal, Supplement
\def\ao{{Appl.~Opt.}}           % Applied Optics
\def\apss{{Ap\&SS}}             % Astrophysics and Space Science
\def\aap{{A\&A}}                % Astronomy and Astrophysics
\def\aapr{{A\&A~Rev.}}          % Astronomy and Astrophysics Reviews
\def\aaps{{A\&AS}}              % Astronomy and Astrophysics, Supplement
\def\azh{{AZh}}                 % Astronomicheskii Zhurnal
\def\baas{{BAAS}}               % Bulletin of the AAS
\def\jrasc{{JRASC}}             % Journal of the RAS of Canada
\def\memras{{MmRAS}}            % Memoirs of the RAS
\def\mnras{{MNRAS}}             % Monthly Notices of the RAS
\def\pra{{Phys.~Rev.~A}}        % Physical Review A: General Physics
\def\prb{{Phys.~Rev.~B}}        % Physical Review B: Solid State
\def\prc{{Phys.~Rev.~C}}        % Physical Review C
\def\prd{{Phys.~Rev.~D}}        % Physical Review D
\def\pre{{Phys.~Rev.~E}}        % Physical Review E
\def\prl{{Phys.~Rev.~Lett.}}    % Physical Review Letters
\def\pasp{{PASP}}               % Publications of the ASP
\def\pasj{{PASJ}}               % Publications of the ASJ
\def\qjras{{QJRAS}}             % Quarterly Journal of the RAS
\def\skytel{{S\&T}}             % Sky and Telescope
\def\solphys{{Sol.~Phys.}}      % Solar Physics
\def\sovast{{Soviet~Ast.}}      % Soviet Astronomy
\def\ssr{{Space~Sci.~Rev.}}     % Space Science Reviews
\def\zap{{ZAp}}                 % Zeitschrift fuer Astrophysik
\def\nat{{Nature}}              % Nature
\def\iaucirc{{IAU~Circ.}}       % IAU Cirulars
\def\aplett{{Astrophys.~Lett.}} % Astrophysics Letters
\def\apspr{{Astrophys.~Space~Phys.~Res.}}
                % Astrophysics Space Physics Research
\def\bain{{Bull.~Astron.~Inst.~Netherlands}} 
                % Bulletin Astronomical Institute of the Netherlands
\def\fcp{{Fund.~Cosmic~Phys.}}  % Fundamental Cosmic Physics
\def\gca{{Geochim.~Cosmochim.~Acta}}   % Geochimica Cosmochimica Acta
\def\grl{{Geophys.~Res.~Lett.}} % Geophysics Research Letters
\def\jcp{{J.~Chem.~Phys.}}      % Journal of Chemical Physics
\def\jgr{{J.~Geophys.~Res.}}    % Journal of Geophysics Research
\def\jqsrt{{J.~Quant.~Spec.~Radiat.~Transf.}}
                % Journal of Quantitiative Spectroscopy and Radiative Transfer
\def\memsai{{Mem.~Soc.~Astron.~Italiana}}
                % Mem. Societa Astronomica Italiana
\def\nphysa{{Nucl.~Phys.~A}}   % Nuclear Physics A
\def\physrep{{Phys.~Rep.}}   % Physics Reports
\def\physscr{{Phys.~Scr}}   % Physica Scripta
\def\planss{{Planet.~Space~Sci.}}   % Planetary Space Science
\def\procspie{{Proc.~SPIE}}   % Proceedings of the SPIE

\let\astap=\aap
\let\apjlett=\apjl
\let\apjsupp=\apjs
\let\applopt=\ao